\documentclass[%
 aip,
 jmp,%
 amsmath,amssymb,
 reprint,%
]{revtex4-2}

\usepackage{graphicx}
\usepackage{dcolumn}
\usepackage{bm}
\usepackage{multirow}

\begin{document}


\title{Origins of Fine Structure in DNA Melting Curves}

\author{Arevik V. Asatryan}
\email{arevik.asatryan@ichph.sci.am}
\affiliation{3D Printing Research Laboratory, A.B. Nalbandyan Institute of Chemical Physics, Yerevan, Armenia}

\author{Albert S. Benight}
\email{abenight@pdx.edu}
\affiliation{Departments of Physics and Chemistry, Portland State University, Oregon, USA}

\author{Artem V. Badasyan*}
\email[ Corresponding author: ]{artem.badasyan@ung.si}
\affiliation{Materials Research Laboratory, University of Nova Gorica, Nova Gorica, Slovenia}

\date{\today}

\begin{abstract}
With the help of one-dimensional random Potts-like model we study the 
origins of fine structure observed on differential melting profiles of 
double-stranded DNA. We assess the effects of sequence arrangement on 
DNA melting curves through the comparison of results for random, 
correlated, and block sequences. Our results re-confirm the  
smearing out the fine structure with the increase of chain length for 
all types of sequence arrangements and suggest fine structure to be a 
finite-size effect. We have found, that the fine 
structure in chains comprised of blocks with the correlation in 
sequence is more persistent, probably, because of increased sequence 
disorder the blocks introduce. 
Many natural DNAs show a well-expressed fine structure of melting profiles. In view 
of our results it might mean the existence of blocks in such 
DNAs. The very observation of fine structure may also mean, that there 
exists an optimal length for natural DNAs \emph{in vivo}. 
\end{abstract}


\maketitle 


\section{\label{sec:intro}Introduction}
Altered external conditions (such as increased temperature) in a system containing double stranded deoxyribonucleic acid (DNA) molecules may trigger a change of their conformation from ordered helical to disordered coil \cite{gros}. This transition is referred to as DNA melting. DNA is a double stranded heteropolymer, comprised of a sugar-phosphate backbone with side groups containing one of four nucleotide bases (cytosine [C], guanine [G], adenine [A] or thymine [T]) \cite{cantor, watson}. Sequence complementarity rules for the double helix require that an A on one of the strands is always paired with T on the other, and G pairs with C. The A--T basepair is stabilized by two hydrogen bonds, and the G--C basepairs have three hydrogen bonds. Due to such compositional disorder, natural DNAs are treated as heteropolymers, and complementarity rules allow to describe DNAs as systems having binary disorder. Each particular sequence of basepairs along the double helix defines the genetic code responsible for the diversity of organisms in nature. Perhaps related to their biological relevance, different sequences give rise to different melting profiles, and display multiple peaks on differential melting curves (DMCs) \cite{gros}. The array of peaks observed along a DMC with increased temperature are referred to as \emph{fine structure} \cite{lyubchenko1976,vartell1985}. 
Water-DNA interactions have been shown to play an important role for DNA conformations both experimentally \cite{lrDNA}, and theoretically \cite{bad11}. How exactly is the fine structure affected by the presence of water, is not clear at this moment, and will be left for future studies.

In the mid-1960s and early 1970s at the beginning of the era of DNA 
melting, theoretical approaches were based on the two-component 
one-dimensional Ising model \cite{azbel1973}  formulated  to predict 
DNA melting curves. When fitted to experimental melting curves, the 
Ising Model approach was able to predict, at least qualitatively,  much 
of the fine structure experimentally observed on DMCs \cite{azbel1979}. 
Perhaps the most successful approaches to predicting DNA DMC 
\cite{lehman1968,frkam1972,vologodskii2018} were based on the 
Poland-Sheraga method \cite{polsher} using the Fixman-Freire 
approximation \cite{fixfr}. These theories were formulated in terms of 
the Zimm-Bragg approach originally applied to analysis of the 
helix-coil transition in polypeptides \cite{zimmbragg,pre10}; but often 
used to describe the conformational changes in DNA as well 
\cite{todd2009,bad2020}. 
The primary assumption underlying these approaches was that nearest-neighbour basepair interactions play a central role in the sequence dependence of DNA melting, and that peaks observed on DMCs correspond to the independent melting of lengthy sequence blocks differing in their percentage of G--C base pairs
\cite{frkam1972,azbel1979}.
For long DNAs where the number of such blocks can be large (several hundred basepairs), the fine structure on DMCs appeared to get smeared due to the overlap of different peaks (see the recent review
\cite{vologodskii2018}). Therefore, it was surmised that clearly 
defined fine structure on DMCs should only be observed for DNA chains 
of moderate length where formation of a small number of loops is allowed.  
 

Unfortunately, this classical viewpoint does not provide insight into 
the relationship between the number of blocks and the number of peaks 
observed on experimentally measured DMCs. While helical segments of a homopolymer have often been
considered as blocks, the question of the range of correlations in 
heteropolymer systems is not a simple one and is closely related to the theory of correlations in disordered systems \cite{cris}. In 2021 the Nobel Prize was awarded to prof. Parisi "for the discovery of the 
interplay of disorder and fluctuations in physical systems from atomic to planetary scales" \cite{parisi}, and our problem belongs to the same category. 

In order to improve the description of many-body effects, and to 
double-check the importance of system length for the appearance of fine structure, another approach has been put forth. It is based on a one-dimensional many-body Hamiltonian with Potts spins \cite{mor1990}. This approach is known as the Generalized Model of Polypeptide Chains (GMPC) and provides an alternative for describing melting of a double stranded DNA homopolymers in the limit of small loops \cite{mor2000}. The GMPC in the thermodynamic limit has been used to study the conformational transitions in both homopolymers \cite{mor1990} and annealed heteropolymers \cite{bad2005,constann}. In order to separately consider the effect of frozen sequence and to depict the origins of the fine structure of DMCs, we have employed the constrained annealing method, which accounts for the existence of two types of degrees of freedom: annealed ones, which can rearrange in order to minimize the free energy and frozen ones, which do not change in time. Thus biopolymer conformations are considered as annealed degrees of freedom, while the sequence of repeating units as frozen \cite{constann,constann191}.As a result of these considerations for infinite chain lengths, it was possible to estimate the melting temperature and melting interval \cite{bad2005,constann,constann191}. The constrained annealing method also resulted in two large peaks on the DMC, but failed in reproducing any of fine structures on DMC \cite{constann,constann191}. Since multi-peak DMCs have been experimentally observed in a number of studies \cite{dalyan,lyubchenko1976,monaselidze2008microcalorimetric}, 
there is an obvious necessity for more in depth theoretical 
investigation on the origins and accurate predictions of fine structure on DMCs of duplex DNAs. 

In the present study, aiming to improve our understanding of 
correlations in helical regions of DNA, we directly generate binary 
random sequences of different finite lengths. The corresponding 
transfer-matrices \cite{bad2005} are directly multiplied to result in 
the partition function, which allows to calculate the degree of 
helicity for each of the generated sequences. Block structures of DNA 
heteropolymers are mimicked by merging sequences with different G--C 
content. Comparison of the results for random, correlated, block random 
and block correlated sequences provides new insights into the origin of 
DMC fine structure.

Specifically, numerical calculations addressed the following questions: 
(i) What is the difference between predicted DMCs for short versus long DNAs? 
(ii) How does the presence of sequence blocks affect DMC fine structure?
(iii) What are the effects of correlations within sequence blocks?

\section{MATERIALS AND METHODS}
In this study we compare the effects of four types of sequence 
schemes onto the DNA melting curves. Considered schemes are: {\bf random, 
correlated, random block, correlated block}.

{\bf Random sequences} with two-component heterogeneity were generated using 
Wolfram Mathematica \cite{WMath}. Each repeating unit (r.u.) comprised 
of G -- C basepair, assigned as type $"A"$, enters the sequence with 
probability $P(A)=x$, and A--T basepair (type $"B"$) with probability 
$P(B)=(1-x)$ \cite{asatryan2018algorithm}. Thus defined, $x$ has the 
usual meaning of G--C fraction. In order to be able to observe strong 
disorder, and not just doping effects, $x$ values from the middle of 
value interval should be taken. In this study two particular values for 
random sequences have been used: $x=0.4$ and $x=0.5$.

\begin{figure}[!ht]
\begin{center}
\includegraphics[width=0.85\textwidth]{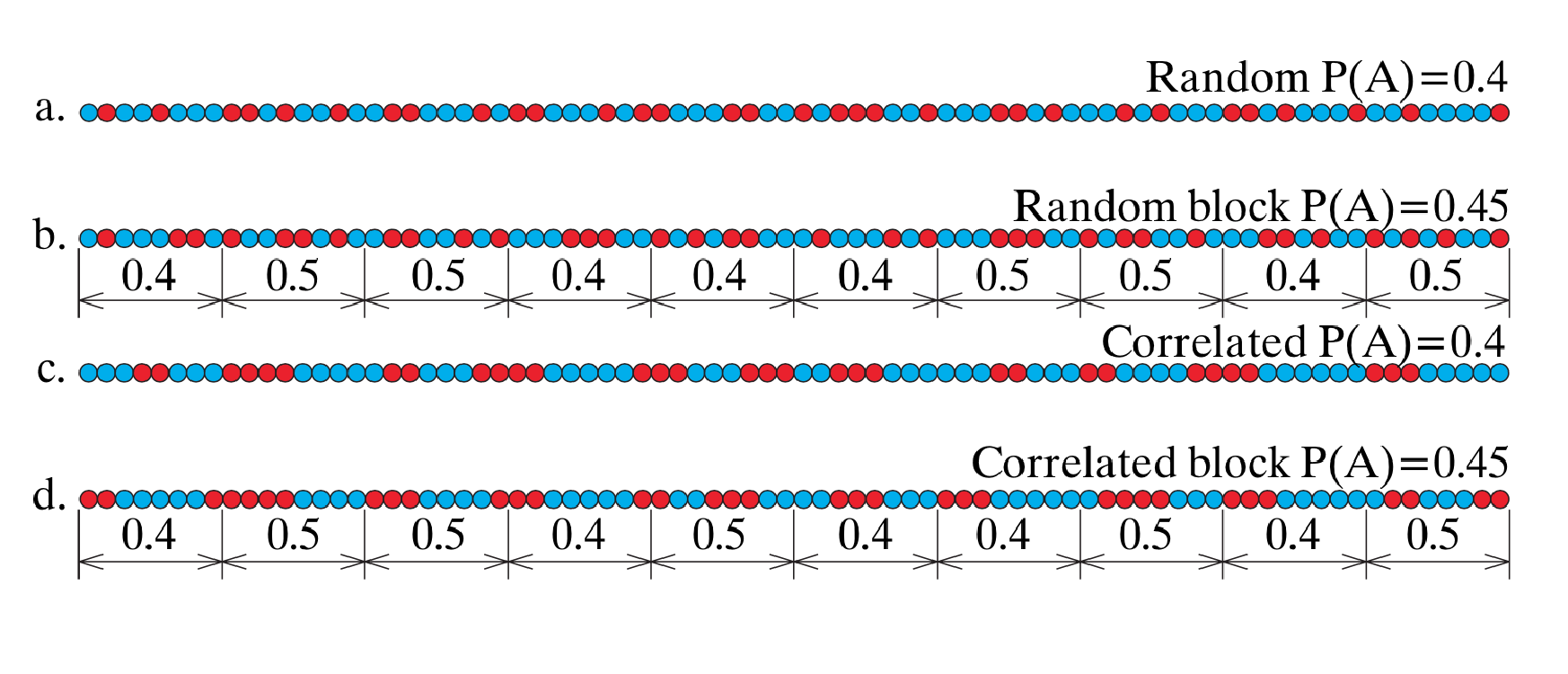}
\end{center}
\caption{Illustration of sequence generation: a) random, b) 
random block, c) correlated, and d) correlated block. Red spheres 
correspond to type $"A"$, and the blue ones to type $"B"$. Block length 
(in the figure, of 8 r.u) is actually 3000-r.u.-long, see text.
}
\label{f0}
\end{figure}

{\bf Correlated sequences} are generated by increasing the probability of
the neighboring basepair of the same type by $\Delta x$, and decreasing the 
probability for different neighboring basepairs by the same $\Delta x$. 
A detailed block-scheme of correlated sequence generation algorithm is 
provided in \ref{algorithm}. Introduction of short-ranged 
correlations in the sequence allows to reveal their effect on melting 
curve fine structure.

{\bf Block sequences} for both random and correlated cases result from randomly merging 
different sequences with different G--C fractions, $x$. Mixing 
(merging) equal amounts of random sequences with $x=0.4$ and $x=0.5$, produces blocked 
sequences with $x=0.45$. Sequence generation is illustrated in Fig.\ref{f0}.

As proposed by Parisi, reduced free energies at fixed disorder content 
($x$, in our case) should tend to a certain limit at infinite system 
sizes. For a given parametrization, the length scale for self-averaging of the free energy 
(not shown) in our model is $N\geq 3000$ r.u. \cite{arev2020,Arevik} 
both for random and correlated sequences. The type of a sequence is not 
affecting the self-averaging, since the self-averaging length scale is 
much larger than the sequence correlation, which affects the 
probability for the nearest neighbour only.
Therefore, in the current study, the temperature dependence of the 
helicity degree is calculated for both random and correlated sequences 
in blocks of 3000 r.u. 

For every sequence, the partition function is given by,
\begin{equation}
\label{part.func.}
Z=\text{Tr} \prod_{i=1}^N G_i,
\end{equation}
\noindent where 
\begin{equation}
\label{trasfermatrix_b}
G_i(\Delta \times \Delta)=
\begin{bmatrix}
e^{\frac{U_i}{T}} &1&0& \cdots &0&0&0\\
0 &0&1& \cdots &0&0&0\\
\cdots &\cdots &\cdots &\cdots &\cdots &\cdots &\cdots &\\
0&0&0 &\cdots & 0&1& 0\\
0&0&0 &\cdots & 0&0& Q_i-1\\
1&1&1 &\cdots & 1&1& Q_i-1 
\end{bmatrix},
\end{equation}
\noindent is the transfer matrix of the GMPC model 
\cite{bad2005} for $i$-th r.u. of either $"A"$ or $"B"$; 
$e^{\frac{U_i}{T}}$ is the energetic parameter where $U_i$ is the 
hydrogen bond formation energy, and $T$ is temperature; $Q_i$ is the number of 
conformations. Dimensions $\Delta$ of the transfer-matrix are determined by the 
number of repeating units, affected by the formation of hydrogen bonds 
in one r.u. and reflects the single strand rigidity. The $\Delta=2$ 
value corresponds to the Zimm-Bragg model \cite{pre10}, applied to 
polypeptides (see \ref{basic model} for the detailed description on correspondence of parameters). To account for a larger rigidity of a single strand DNA, we use $\Delta=4$ value throughout this study. 

As opposed to treating the sequence disorder in partition function using approximate methods 
\cite{bad2005,constann}, applicable for the infinite chain length $N$, 
we use the straightforward multiplication of matrices for each 
generated sequence, which is exact and valid for any chain length. Also, the approach can be extended beyond 
the calculation of melting temperature $T_m$ and melting interval $\Delta 
T$. That is, the degree of helicity $\theta$ (fraction of intact basepairs) can be determined according to: 

\begin{equation}
\theta=\frac{\text{Tr}[E,O]\prod_{i=1}^{N}\hat{M_i}\left[
\begin{array}{ccc}
O \\
E \\
\end{array}
\right]
}{N\text{Tr}[E,O]\prod_{i=1}^{N}\hat{M_i}\left[
\begin{array}{ccc}
E \\
O
\end{array}
\right]
}; \\ 
\hat{M}_i=\left(
\begin{array}{ccc}
\hat{G}_i&\hat{G}_i^\prime\\
O&\hat{G}_i
\end{array}
\right).
\label{theta}
\end{equation}
\noindent where $E$ is the unit and $O$ is null matrix and $\hat{M}_i$ 
is the supermatrix.
Here, $\hat{G}^\prime_i=\partial \hat{G}_i/ \partial J_0$, so the first 
element of the transfer matrix $\hat{G}^\prime_{i, 11}=e^{J_i}$ and all 
other elements are zero. For the details of the model 
definition please consult the \ref{helicity degree}.  

Calculations are performed by multiplication of transfer-matrices in 
Eq. \eqref{part.func.} for each type of generated sequence. From 
Eq. \eqref{theta} each 
supermatrix depends on the type of r.u., and the degree of helicity 
represents every particular sequence and is unique. 

On a Macintosh HD running MacOS 14.1.1 with 16 GB of memory and an Apple 
M1 Pro chip, the process of calculating the helicity degree and gathering 
tensor data for each DNA sequence consisting of 3000 repeated units requires 
approximately 2.5 hours. Subsequently, the collected data is utilized for 
merging sequences, and the merging process along with helicity degree calculation 
varies in duration, ranging from several seconds to 3 minutes. The time 
required depends on the number of sequences being merged.

DMCs are obtained from the numeric derivation of Eq.~\eqref{theta}. The 
estimated numerical error is of the order of $10^{-5}$.

\section{RESULTS AND DISCUSSION}

\subsection{Dependence of the DMC profile on length, block structure and 
sequence correlations}

\begin{figure}[!ht] 
\includegraphics[width=0.85\textwidth]{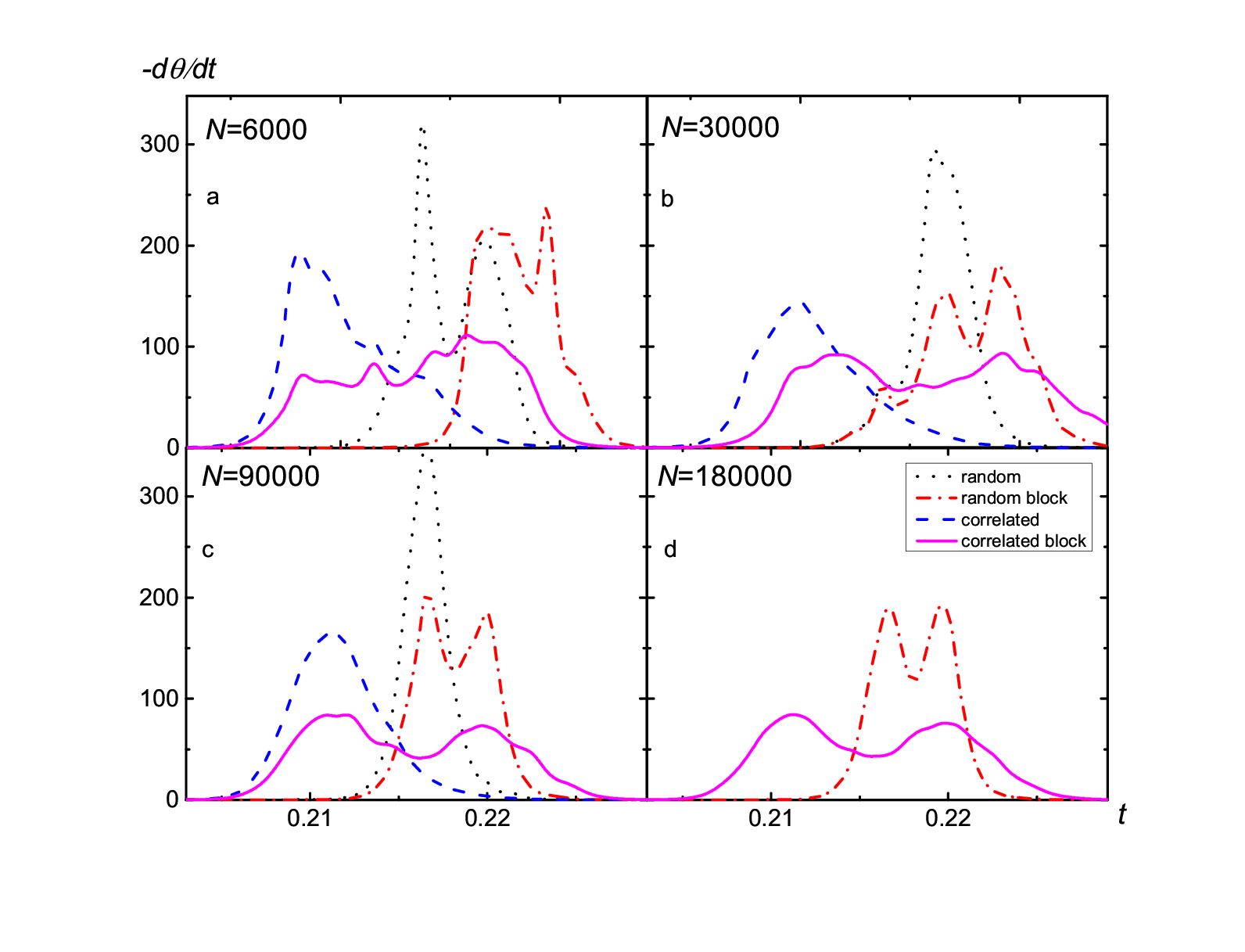}
\caption{\label{DMCN} Calculated DMCs for long heteropolymers with $N=6000$ (a), $N=30000$ (b), $N=90000$ (c) and 
$N=180000$ (d). Curves with $x=0.4$ are compared 
with those for the joint block scheme generated with 
parameters $x=0.4$, $x=0.5$, $\Delta x=0.3$. The other parameters, same 
for all curves, read: $U_A=1$, $U_B=0.8$, $Q_A=71$, 
$Q_B=51$, $\Delta =4$. Temperature is given in units of $t=T/U_A$. 
The legend indicates the specific sequence arrangements.}
\end{figure}

Equation~\eqref{theta} enables the examination of sequence dependent 
features such as length, block structure and sequence correlations, on 
the DMC. Obtained curves visually remind those familiar from the experiments. 
Thus, the dashed blue line of Fig.\ref{DMCN}a 
($N=6000$, correlated sequence) displays a profile visually similar to the experimental DMC of 
calf thymus DNA \cite{dalyan,monaselidze2008microcalorimetric}.
Since each particular DMC depends on sequence, different 
sequences are incomparable, and only general trends can be deduced from 
results of the calculations. 

As seen in Fig.\ref{DMCN}, fine structure is present at shorter 
sequence lengths for all sequence arrangements considered (random, random block, 
correlated and correlated block), and DMC turns into a smooth curve, 
when sequences become longer, a trend first reported by 
Lyubchenko \emph{et al} \cite{lyubchenko1976}. 

In absence of a block structure, random and correlated sequences 
(dotted and dashed lines, Fig.\ref{DMCN}) result in a smooth curve with a 
single maximum, while random block and correlated block sequences 
(dash-dotted and solid lines, Fig.\ref{DMCN}) give rise to a smooth 
curve with two well-expressed and wide maxima. Most persistent in terms of a 
fine structure is the correlated block scheme. 

It is tempting to suggest an explanation for the small peaks forming 
the DMC profile as appearing from individually melting regions, following Lyubchenko \emph{et al} 
\cite{lyubchenko1976}. Using the expression for the transition interval 
of melting of a homopolymer, the same authors have qualitatively estimated 
the area of a peak of fine structure to be proportional to $\nu/N$ of the total area of the differential
melting curve ($\nu$ is the average length of a helical region). The 
larger is the chain length $N$, the smaller will be the contribution 
from the peak, and eventually, peaks disappear for large $N$. 
The length of an individually melting region should be of the order of a spatial 
correlation length, which formally brings the problem to the 
calculation of the correlation function, which is not a trivial 
question for disordered systems, heteropolymer, in this case. Probably, 
this is the reason, why Lyubchenko \emph{et al} 
\cite{lyubchenko1976} have to rely on the transition interval formula 
for a homopolymer. As it was reported before\cite{gros,bad2005}, the span of correlations within the secondary structure of a 
heteropolymer is certainly different from that of a homopolymer. This is 
obvious from the fact that the transition interval of a heteropolymer has a different functional 
dependence on model parameters than a homopolymer. Thus, while valuable as an 
idea, the concept of Lyubchenko \emph{et al} needs to be justified, 
when it comes to the scale of conformational correlations for the 
heteropolymer. 

With a very interesting numerical experiment, Lyubchenko \emph{et al} have illustrated the relevance of 
the value of nucleation parameter $\sigma$ for the fine structure. Fig.1 
of Ref.~\cite{lyubchenko1976} shows a calculated DMC for $N=30000$ r.u. 
long random heteropolymer in a model that allows for loop formation, 
and apparently results in fine structure. Fig.3 of 
Ref.~\cite{lyubchenko1976} is for exactly the same parameters, but 
without loops, and one can see the fine structure disappeared. What has 
changed? The presence of loops introduces additional cooperativity into 
the system, and significantly decreases the value of nucleation 
parameter $\sigma$. When the loop factor is removed from the model, the 
cooperativity decreases, and the value of $\sigma$ increases. So, which 
factor is relevant {\it per se}, loop formation possibility, or the 
value of the nucleation parameter $\sigma$? Lyubchenko \emph{et al} 
gave a clear answer to the question in Fig.4 of 
Ref.~\cite{lyubchenko1976}, where they took the model without loops, 
but decreased the value of $\sigma$ to the value it had in the case of 
loops, and have demonstrated the re-appearance of fine structure. So, 
according to Lyubchenko \emph{et al},
it is all about the values of $\sigma$: at small values there is a 
fine structure, while at large ones there are no signs of fine 
structure. If recall the simple relationship between the parameter 
$\sigma$ and the correlation length $\xi$, $\sigma=\xi _{max}^{-2}$ provided 
by the homopolymer version of the GMPC model (please see 
Eq. (S3) of SI and references therein), one can re-read the message: 
large conformational correlations (small $\sigma$s) result in fine 
structures. Similar to the preceding paragraph, here again, the need to estimate the span of conformations 
for a heteropolymer becomes apparent. Let us say it clear, although we 
do not study the correlation function for the heteropolymer in this 
paper, the work of Lyubchenko \emph{et al} inspires a clear working 
ansatz for the future study: fine structure is a finite size effect, 
related to the $\xi/N$ ratio. When the ratio is comparable with unity, finite size 
effects, including the fine structure, are apparent, while at small
values the system self-averages nicely, so that no fine structure is 
present.

As we illustrate in Fig.\ref{DMCN}, besides the 
abovementioned factors, the specific sequence arrangements 
can also seriously affect the DMC shapes. That is, our 
calculated DMCs of random sequences are smooth for $N=30000$ 
(Fig.\ref{DMCN}b, dotted), while the presence of block structure 
results in fine structure (Fig.\ref{DMCN}b, dash-dotted). 
The significant difference between our approach and that of Lyubchenko \emph{et al} is 
our utilization of sequence blocks. Pieces of random sequence of 3000 
nucleotides each were linked to obtain the block structure in our 
study. In contrast,  Lyubchenko \emph{et al} generated random 
sequences, and considered chain pieces of one helical segment long as 
individual blocks. 

To conclude this section, results presented in Fig.\ref{DMCN}, 
qualitatively support the view of fine structure as a finite size 
effect, and the expression of it depends on the disorder, encoded in the different schemes of sequence organization.

\subsection{Effects of averaging over the sequences}
DMCs for different sequences generated at the same value of $x$ mimic 
melting curves for different DNA sequences with the same G--C content. 
The theory of systems with a random 
potential, one example being the model of heterogeneous sequence DNA 
melting\cite{bad2005}, similar to that considered here, provides 
estimates of quantities, averaged over the disorder, in the limit of 
infinite system size \cite{cris}. Understanding the mechanisms behind 
this averaging may illustrate effects on the DMC, arising 
from particular sequence structure of DNA.

\begin{figure}[!ht]
\includegraphics[width=0.85\textwidth]{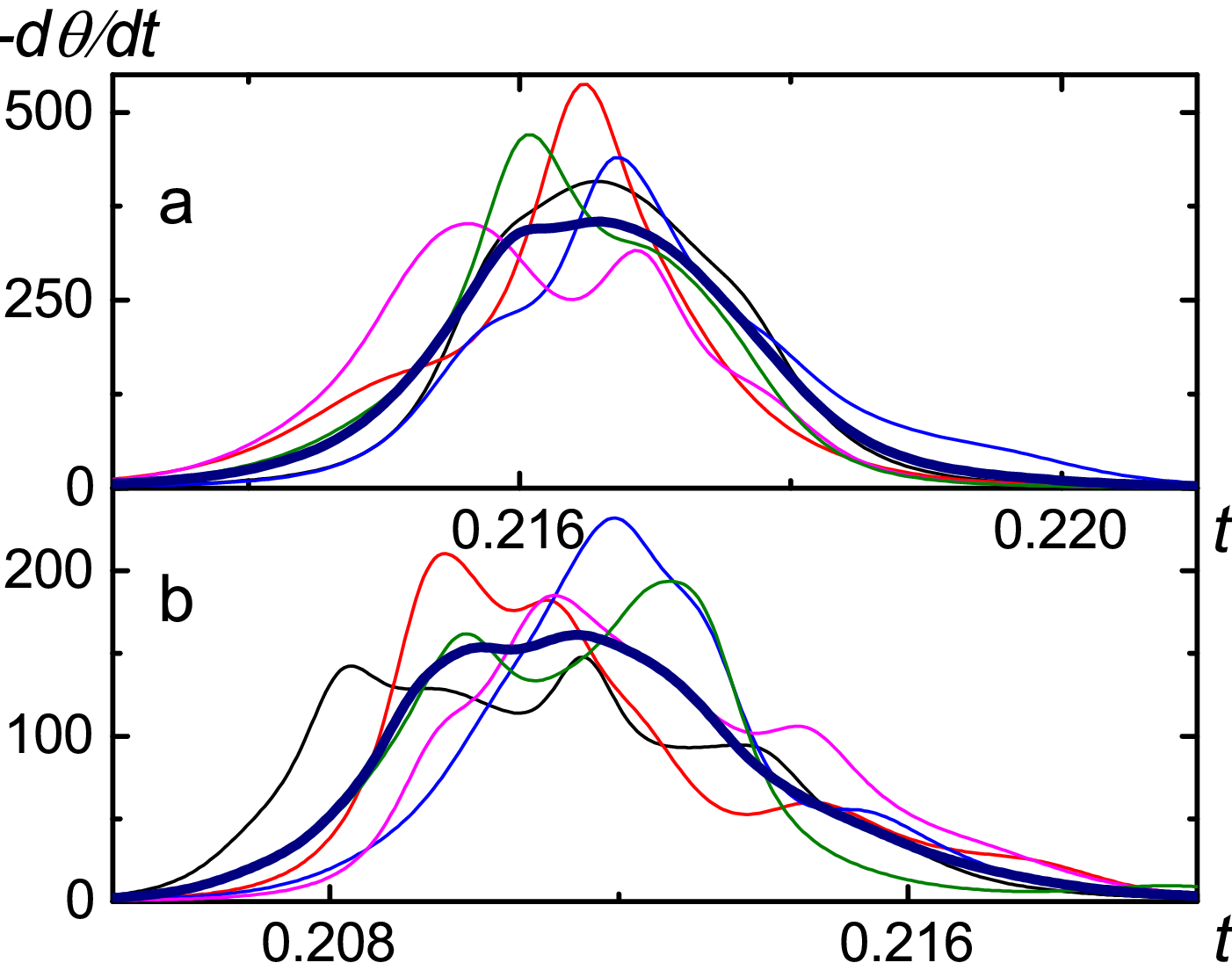}
\caption{\label{f.dmc.initial_average} Calculated DMCs for heteropolymers (colored 
thin lines) and their average (thick line). (a) random sequences: (b) 
sequences with correlation. $x=0.4$, other parameters are as in Fig. 
\ref{DMCN}. Averages were determined from 10 individually calculated curves (not all shown).}
\end{figure}

We start by generating ten sequences with $N=3000$ at fixed $x$, 
calculating the DMC and then averaging the curves. As 
shown in Fig.\ref{f.dmc.initial_average}, the curves obtained do not 
follow a pattern. Some have well-defined fine structure, others do not, 
however, the fine structure is more expressed on DMCs for correlated 
sequences (see more curves in \ref{all_DMC} of \ref{melting par}). 
Interestingly, for both the uncorrelated 
(Fig.\ref{f.dmc.initial_average}a) and correlated 
(Fig.\ref{f.dmc.initial_average}b) sequences, results of averaging are 
similar: smooth curves with two weakly expressed maxima are obtained.  

\subsection{Comparison between block-averaging and sequence-averaging}

\begin{figure}[!ht]
\includegraphics[width=0.85\textwidth]{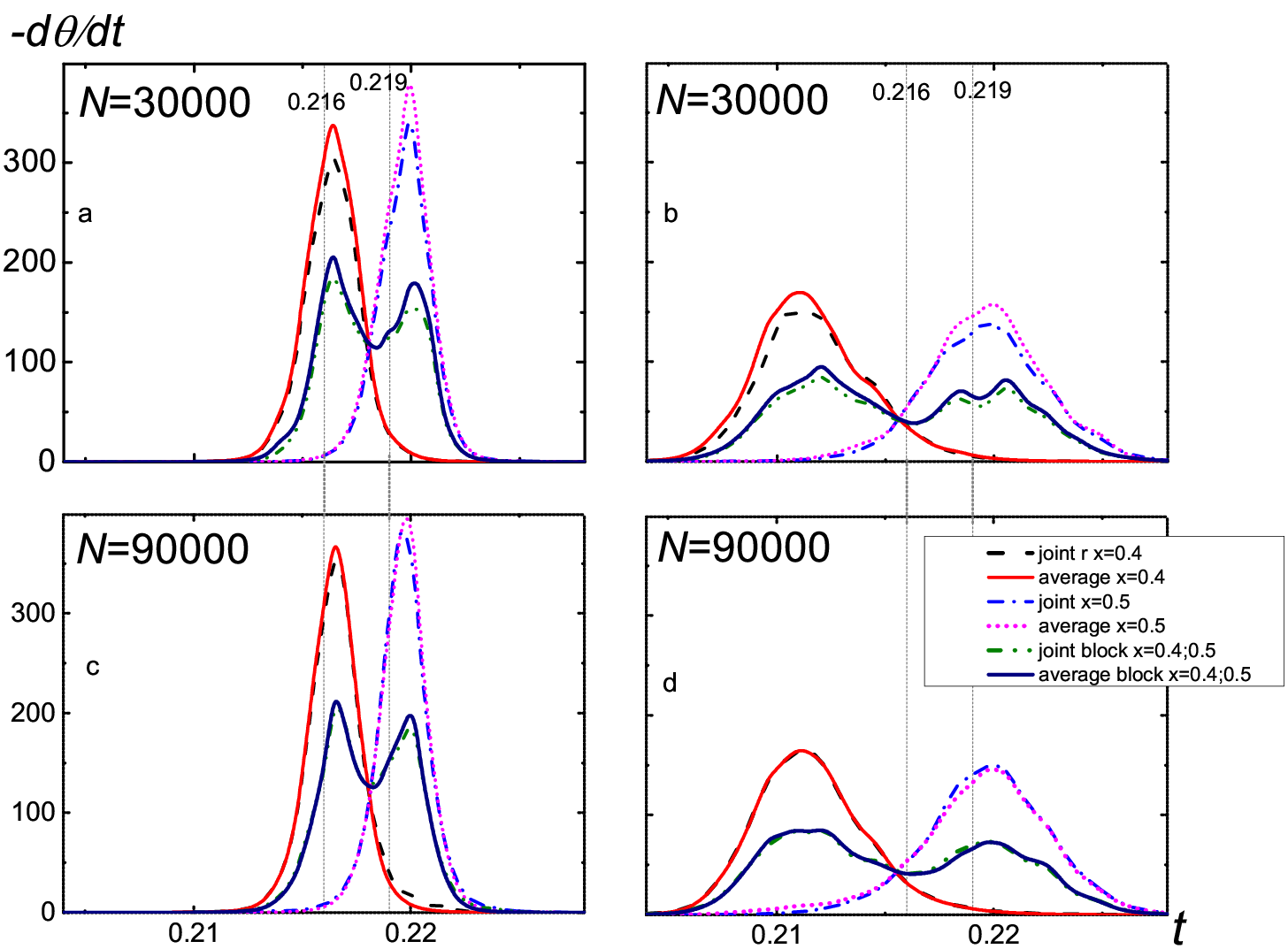}
\caption{\label{f.DMC.block.av} Joint DMC of heteropolymers in 
comparison with corresponding average DMC: (a), (c) random sequences, 
(b), (d) sequences with correlation. The parameters are as in Fig. 
\ref{DMCN}. Melting temperatures calculated from $T_m=xT_A+(1-x)T_b$ are indicated for joint 
heteropolymers as dashed thin vertical lines.}
\end{figure}

Results in Fig.\ref{f.dmc.initial_average} show, that averaging over 
large number of shorter sequences results in a smooth curve, in a way 
similar to the long sequence behavior. To check if it is true, we take 
10 (30) sequences, $N=3000$ r.u. each, calculate the average DMC (codename 
{\bf average}), then compare it with the DMC of a {\bf joint} 
sequence, made by gluing together exactly the same 10 (30) sequences. 
Both the random and correlated curves (Figs.~ \ref{f.DMC.block.av}a, c 
and \ref{f.DMC.block.av}b, d respectively) look quite similar, with a minimal number of peaks. 
The tendency to a shapeless DMC without fine structure is obvious. 
This demonstrates qualitatively similar DMCs result for sequences, 
made by gluing blocks into a large single chain and for those obtained by 
averaging over a large number of independent blocks. 
More joint curves are shown on \ref{all_merged} in \ref{melting par}. Thus 
the {\it ansatz} is confirmed: whether shorter sequences are joint into 
a long chain, or results for short sequences are averaged, resulting 
DMC is same. And longer is the chain length (or larger is the number of 
shorter sequences), the better is agreement between the two (compare 
Figs.~ \ref{f.DMC.block.av}a,b with Figs.~ \ref{f.DMC.block.av}c,d).

Dashed thin vertical lines on Fig.~\ref{f.DMC.block.av} indicate the 
melting temperatures for random sequences, drawn according to the 
theoretically calculated value \cite{Artyom, bad2005} for infinitely long 
sequences as $T_m=xT_A+(1-x)T_b$ (for $x=0.4$ and $x=0.5$ values). For 
random sequences, Fig.~\ref{f.DMC.block.av}a,c, the maxima of DMCs are 
very close to the corresponding random heteropolymer values (dashed 
vertical lines), while the introduction of block structure in 
Fig.~\ref{f.DMC.block.av}b,d, breaks the agreement with the theoretical 
values. 

\begin{figure}[!ht]
\includegraphics[width=0.85\textwidth]{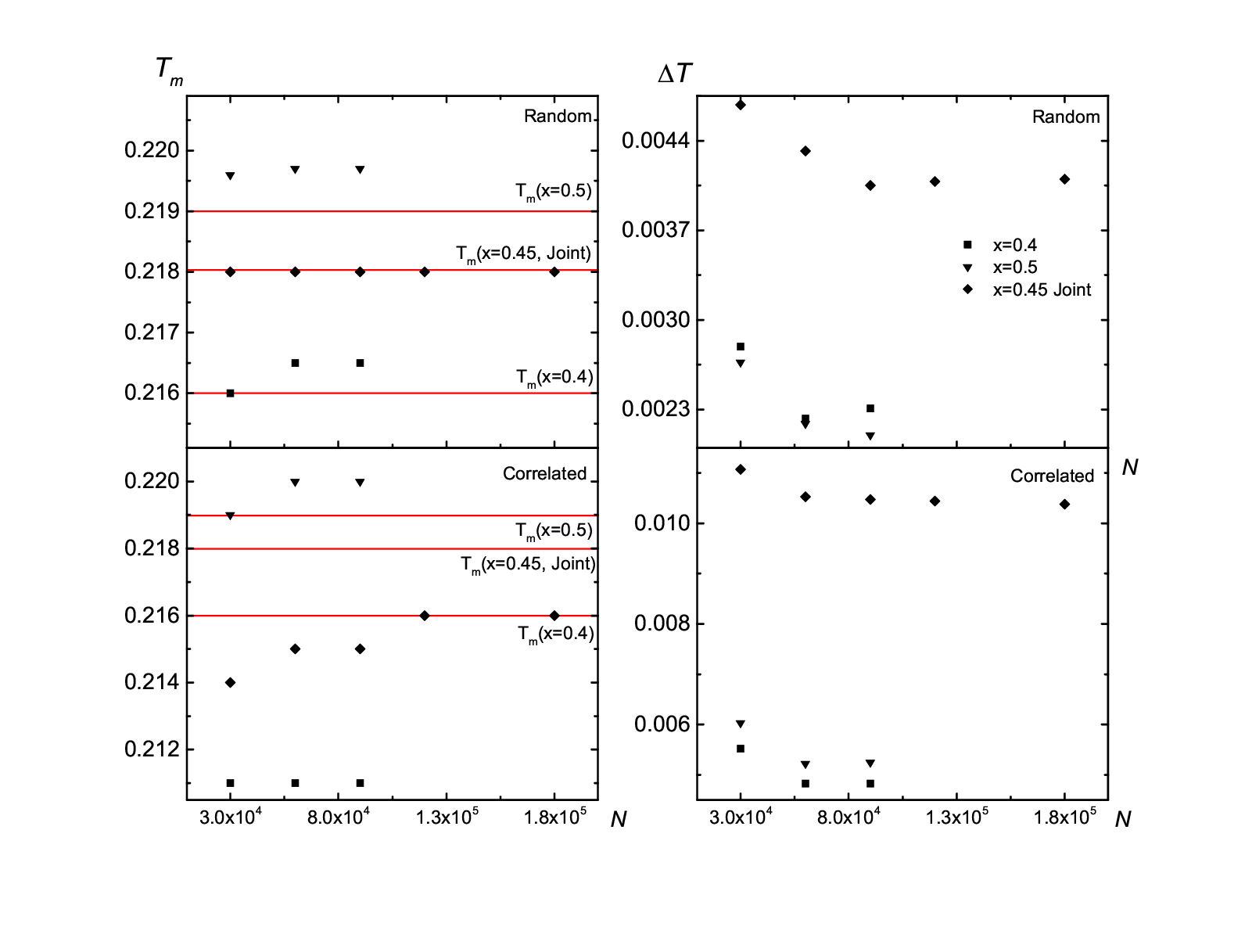}
\caption{\label{Tm-N} The dependence of melting temperature (left) and 
melting interval (right) on the sequence length. The red line shows the theoretical value (see 
Ref.~\cite{bad2005}).} 
\end{figure}

To study this question further, we plot the positions of maxima 
(considered as melting temperature) and transition interval vs the 
chain length in Fig. \ref{Tm-N} (left). One can see that for random sequences 
the melting temperature is close to the theoretical value and for the 
correlated heteropolymers there is no agreement between the theoretical 
melting temperatures. As to the melting interval (Fig. \ref{Tm-N}, 
right), it is also strongly affected by the sequence organization. 
Thus, when the random and block sequences are compared, the melting 
interval of the second is about twice bigger. When the random and 
correlated sequences are compared, the melting interval is again twice 
bigger for correlated sequences. And finally, when the block sequences 
are compared with correlated block ones, the melting interval is again 
approximately twice bigger, thus it is about four times bigger than 
melting interval of random sequences. Additional info on melting 
temeperatures and intervals can be found in Table 1 of \ref{helicity degree}.


\section{Summary and conclusions}
Our results indicate, that sequence organization strongly affects 
the presence or absence of fine structure (Fig.\ref{DMCN}). When 
different sequence schemes are compared, the expression of a fine 
structure (at fixed sequence length $N$) increases in the following 
order: {\bf random $\to$ correlated $\to$ random block $\to$ random 
correlated}. 

The fact, that increasing sequence length smears out the fine structure, means its presence on 
DMC curves is a finite-size effect, which depends on a ratio $\xi/N$, 
where the spatial scale $\xi$ should be related to the correlation 
length of the system. The calculation of the correlation length of a 
system with disorder (heteropolymer) is related to the calculation of 
the second Lyapunov exponent\cite{cris}, and is not a trivial task {\it 
per se}. 

Introduction of blocks should decrease the physical cooperation between 
different parts of the system due to the additional disorder it 
creates as compared to the random sequence. The better expression of fine structure (see Fig.\ref{DMCN}) 
for the block sequences (at same $N$) can be understood as decreased value of the correlation 
length $\xi$ of the system. However, since the calculation of the correlation length has not been done here, it is a speculation, for the moment.

Results of Fig. \ref{DMCN} showed that random and correlated sequences 
give rise to a single peak DMC curve at large $N$, while the sequences 
with blocks of different G--C content $x$ result in two peaks. This confirms our earlier results regarding the presence of just two peaks, 
obtained using the method of constrained annealing (CA) \cite{constann191,constann}. 
Based on this observation we claim that the CA method is a good approximation for completely random heteropolymers. 
We notice correlated sequences resulting in more peaks on the DMCs than 
uncorrelated sequences. The number of peaks on the DMC tend to increase 
for block sequence structure. We also see the tendency to smear out all the 
peaks of fine structure at increased sequence length, in agreement with 
Lyubchenko et al \cite{lyubchenko1976}. The very fact, that many DNAs of 
living organisms show the presence of fine structure on their melting 
profiles, may be considered as a sign, that they are optimized not to 
exceed a certain length, the biological meaning of which has still to be 
clarified. 

Our study suggests insights into the stability of DNA based on its primary 
structure and length. Targeted gene therapy crucially depends on DNA conformations 
to avoid biodegradaton during the delivery phase \cite{dd1,dd2}. It makes our findings 
relevant for future development of drugs intended for use in gene therapy. 

\section*{Conflict of Interest Statement}

The authors declare that the research was conducted in the absence of any commercial or financial relationships that could be construed as a potential conflict of interest.


\section*{Funding}
A.A. acknowledges funding of Science committee RA in frame of scientific project 19YR-1F057 and the scientific project N22rl-012.
A.B. acknowledges the partial financial support from Erasmus+ Project 
No. (2023-1-SI-KA171-HED-000122882).

\begin{acknowledgments}
The authors would like to thank dr. V. Morozov and dr. Y. Mamasakhlisov for valuable discussions and advices.
\end{acknowledgments}

\section*{Supplemental Data}
 
\section*{Data Availability Statement}
The datasets [GENERATED/ANALYZED] for this study can be found in the GitHub repository: https://github.com/AsatryanArevik/Publication-data.

\appendix

\section{The basic model for heteropolymers}
\label{basic model}

A microscopic Potts-like one-dimensional model with $\delta$-particle interactions 
is used to describe the helix-coil transition in polypeptides from 
early 1990s \cite{mor1990,mor1995}. Later it was shown that the same 
approach could be applied to DNA if the large-scale loop factor
is ignored \cite{mor2000}. This model, known as the Generalized Model
of Polypeptide Chains (GMPC) has the Hamiltonian of the form:

\begin{equation}
\label{hamiltonian}
-\beta \mathcal{H}=J\sum_{i=1}^{N}\delta _i^{(\Delta)},
\end{equation}
\noindent 
where the summation is performed across all $N$ repeated units,  
$\beta=1/T$ is inverse temperature, $J=U/T$, $U$ is the hydrogen bond 
formation energy, and 
$\delta_l^{(\Delta)}=\prod_{k=0}^{k-1}\delta(\gamma_l, 1)$, where 
$\delta(x,1)$ is the Kronecker symbol. $\gamma_l$ is a Potts spin, 
describing the conformational state of repeated unit, that can vary 
from 1 to $Q$ values. $\gamma_l=1$ is taken as helical state and all 
the other values of $\gamma$ correspond to coil states. Thus, the 
presence of the Kronecker delta in the Hamiltonian guarantees that the 
energy $J$ results only when all $\Delta$ adjacent repeated units are 
in the helical conformation. Consequently, this considers the 
limitations on backbone chain conformations imposed by the formation of 
hydrogen bonds. The transfer-matrix, corresponding to the Hamiltonian 
(\ref{hamiltonian}) reads \cite{mor1990}:
\begin{equation}
\label{trasfermatrix_bb}
G(\Delta \times \Delta)=
\begin{bmatrix}
W &1&0& \cdots &0&0&0\\
0 &0&1& \cdots &0&0&0\\
\cdots &\cdots &\cdots &\cdots &\cdots &\cdots &\cdots &\\
0&0&0 &\cdots & 0&1& 0\\
0&0&0 &\cdots & 0&0& Q\\
1&1&1 &\cdots & 1&1& Q 
\end{bmatrix},
\end{equation}
\noindent where $W=e^{\frac{U}{T}}$, has the characteristic equation 
$\lambda^{\Delta-1}(\lambda-W)(\lambda-Q)=(W-1)(Q-1)$. Once the 
eigenvalues are found, the partition function of the model can be 
estimated (at large $N$) as $Z=\lambda_1^N$, and the spatial 
correlation length as $\xi= \ln^{-1}(\lambda_1/ \lambda_2)$, where 
$\lambda_1, \; \lambda_2$ are the maximal and the second largest 
eigenvalues, correspondingly. The point of closest approach of two 
largest eigenvalues determines the transition point estimated from $W 
\approx Q$ as $T_m=U/{\ln Q}$. The correlation length is the distance, 
over which the conformations of repeated units are correlated. 
Since the range of interactions is finite ($\Delta<\infty$), the 
conformational transition in this 1D system is not a phase transition 
(Pierls-Landau theorem)\cite{landau}. Therefore, the correlation length does not 
tend to infinity and remains finite, but reaches its maximum 
$\xi_{max}\propto Q^{\frac{\Delta-1}{2}}$ at the 
transition point \cite{mor1990,bad2013}. Nucleation parameter $\sigma$ and equilibrium constant $s$ from 
Zimm-Bragg model are in correspondence with the following parameters in 
GMPC \cite{mor1995}:

\begin{equation}
\label{cor_ZB}
\sigma=\xi _{max}^{-2}, \quad s=\frac{W}{Q}.
\end{equation}

However, one should consider, that above results are derived for a 
homopolymer model, and are certainly different for a heteropolymer GMPC. 
DNA is a heteropolymer since the adenine-thymidine $(A-T)$ and 
guanine-cytosine $(G-C)$ base pairs differ in the number of hydrogen 
bonds. We model the structural disorder of DNA through the dependence 
of $W$ parameters on the position of r.u. in the sequence. Thus, the 
partition function for a given sequence of base pairs can be written as
\begin{equation}
\label{part.func.A}
Z=\text{Tr} \prod_{i=1}^N G_i,
\end{equation}
\noindent  where the transfer matrix Eq.~\eqref{trasfermatrix_bb} is 
modified to read
\[ G_i= 
\begin{cases}
 G_{AT}  \text{ if } i \text{ -th  r.u. is of }  $A-T$  \text{ type}\\
 G_{GC}  \text{ if } i \text{ -th  r.u. is of }  $G-C$  \text{ type} .
\end{cases}
\]
Given that these matrices are not commutative, every sequence within 
the collection of sequences of length $N$ and disorder concentration 
$x$ possesses distinct statistical characteristics. Typically, each 
specific chain can be identified by a sequence-dependent free energy 
$F_{seq}$. Nevertheless, it is widely accepted that the free energy 
adheres to the principle of self-averaging. This principle asserts that 
the probability distribution of free energies for independent samples 
is highly constrained, resulting in the virtual alignment of the free 
energy with the mean free energy for nearly all sequences. This 
self-averaging tendency has been shown to come to saturation in the 
range of $2000-3000$ r.u. length of the sequence depending on the 
peculiarities of sequence generation \cite{Arevik}. In the past, we 
have treated general characteristics of such heteropolymer model, using 
the microcanonical \cite{bad2005} and the constrained annealing 
\cite{constann191} methods. The essence of a microcanonical method is 
that the quenched averages can be substituted by the annealed average 
at fixed disorder concentration $x$, since the two quantities are equal to 
each other up to the fluctuations of $x$. The constrained annealing is 
a more sensitive method, and treats the conformations as annealed 
degrees of freedom, while the sequences of r.u. is considered frozen. 
Interested reader is addressed to Refs.~\cite{bad2005,constann191}, and 
references therein. Using these approaches the expressions for the 
transition temperature 
\begin{equation}
\label{tm}
T_m = x T_{GC} + (1-x) T_{AT},
\end{equation}
\noindent where $T_{GC}$ and $T_{AT}$ are corresponding homopolymer 
melting temperatures, and the interval 
\begin{equation}
\label{dt}
\Delta T = 2x(1-x) \ln Q (T_{GC} - T_{AT})^2/T_m
\end{equation}
\noindent have been found. None of the approaches gave access to fine 
structure.

\section{Algorithm of correlated sequence generation}
\label{algorithm}

\begin{figure}[ht!]
\centering
\includegraphics[width=0.85\textwidth]{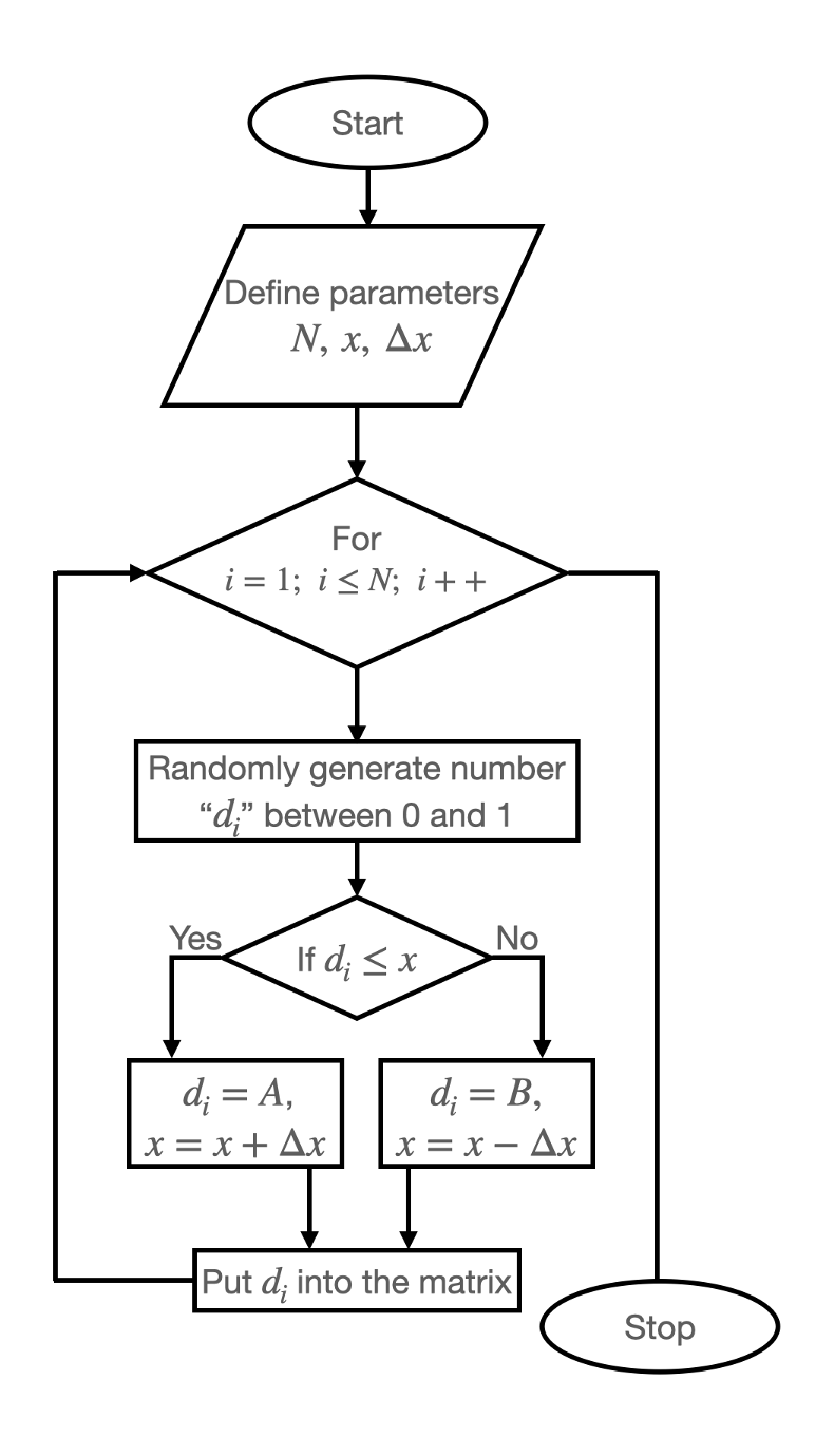}
\caption{\label{blockscheme} Block-scheme for correlated sequence generation}
\end{figure}

When the parameters are defined: $x=0.4$ (G-C fraction), $\Delta x=0.3$ correlation parameter, and $N=3000$ number of repeated units, there is random generation of a number between $0$ and $1$. The random number is compared with $"x"$ and if it is smaller or equal to $"x"$ then, sequence is filled with $"A"$ type R.U. (corresponding to G-C bp) and in the same step $"x"$ is increased with amount of  $\Delta x$ to increase the probability of the next R.U. being generated as  $"A"$ again, if no, the sequence is filled with  $"b"$ type R.U. (A-T bp) and $"x"$ is decreased with amount of  $\Delta x$ to increase the probability of the next R.U. being generated as  $"B"$ again. The value of  $\Delta x$ is chosen comparable to the value of  $"x"$, so that the correlation is noticeable in the sequence and influences the DMCs. 

\pagebreak

\section{Detailed calculation of helicity degree for heteropolymer}
\label{helicity degree}

The Hamiltonian for a heteropolymer in frame of GMPC is as follows:

\begin{equation}
\label{3.2.1}
-\beta H=\sum_{i=1}^{N}J_i\delta_i^{(\Delta)}.
\end{equation}

Where $J_i=\frac{U_i}{T}$ and the energy of hydrogen bond $U_i$ depends on the type of repeated unit, $\delta_j^{(\Delta)}=\prod_{k=\Delta-1}^0\delta(\gamma_{j-k},1)$, where $\gamma_i=1,2,...,Q_i$. In this paper the model of DNA is taken as a sequence of repeated units with bimodal heterogeneity both by energies of helicity structure formation, and by number of conformations of repeated units. Thus a variable $\sigma_i$ is introduced, to take value $1$ with given probability $x$ and $-1$ with probability $(1-x)$. Accordingly, $x$ is the fraction of type $A$ repeated unit in the system: $x_A=\frac{N_A}{N_A+N_B}$. Therefore, we deal with two-component heteropolymer and the intramolecular hydrogen bond's energies of $A$ and $B$ types of repeated units can be expressed as: $J_i=J_0+\Delta J\sigma$ $$J_A=J_0+\Delta J  \text{  and  }  J_B=J_0-\Delta J.$$

Where $J_0$ is energetic parameter to be changed accordingly with the type of repeated unit. According to \cite{mor2000, Artyom, flory}, 
partition function determines as 

\begin{equation}
\label{3.3.0}
Z=J^*\prod G_i J \quad \text{where} \quad
 J^*=
\left(
\begin{array}{ccccccc}
0 & 0 & 0 & \cdots & 0 &1
\end{array}
\right),
\quad
J=
\left(
\begin{array}{ccccccc}
0\\0\\0\\ \cdots \\0\\1
\end{array}
\right).
\end{equation}

However, it has been shown \cite{Arevik} that for a heteropolymer longer than $30$ nucleotides, eq. \ref{3.3.0} can be replaced  with the following one with high precision:

\begin{equation}
\label{3.3.1}
Z=Tr\prod_{i=1}^{N}{G}_i,
\end{equation}



While calculating the partition function according to Eq.\eqref{3.3.1}) for long biopolymers, sometimes the values exceed the limits of the technical support (Wolfram mathematica). To avoid such disturbances we made the following transfiguration:

\begin{equation}
\label{3.3.1.0}
G_i=\lambda_{1i}g_i,
\end{equation}

where $\lambda_{1i}$ is the principal eigenvalue for the transfer-matrix $G_i$. Thus, partition function for bimodal heterogeneity will appear as:
\begin{equation}
\label{3.3.1.1}
Z=\lambda_{1A}^x\lambda_{1B}^{1-x}Tr\prod_{i=1}^N g_i.
\end{equation}

Where $\lambda_{1A}$ and $\lambda_{1B}$ are principal eigenvalues of type $A$ and type $B$ repeated units correspondingly. 

It is generally known, that the following expression can be used to calculate helicity degree of the system: 

\begin{equation}
\label{3.3.3}
\theta_N=\frac{N_h}{N}=\frac{1}{NZ}\sum_{i}\delta_i^{(\Delta)}\exp^{-\beta H}.
\end{equation}

Where $N_h$ is the number of repeated units in helical state.
Taking into account the Hamiltonian expression ( \ref{3.2.1} ), for helicity degree we obtained:

\begin{equation}
\label{3.3.4}
\theta_N=\frac{1}{NZ}\frac{\partial Z}{\partial J_0}.
\end{equation}

With the help of partition function expression in terms of transfer-matrix, we will obtain:

\begin{equation}
\label{3.3.5}
\theta_N=\frac{1}{NZ}\sum_{i}Tr\prod_{k=1}^{i-1}G_kG_i^{\prime}\prod_{k=i+1}^{N}G_k.
\end{equation}

For calculation of the latest equation we have used method of supermatrices. The supermatrix we have inserted has dimensions($2\Delta\times 2\Delta$) and can be expressed as:

\begin{equation}
\label{3.3.6}
\hat{M_i}=\left(
\begin{array}{ccc}
\hat{G_i}&\hat{G_i^\prime}\\
O&\hat{G_i}
\end{array}
\right),
\end{equation}

where O is null matrix ($\Delta\times\Delta$). 
\begin{equation}
\label{3.3.6.1}
\hat{G_i^\prime}=\frac{\partial\hat{G_i}}{\partial J_0},
\end{equation} 
consequently, the matrix has only one nonzero element: $\hat{G}_{i,11}^\prime=\exp^{J_i}$, all the other elements of matrix are equal to zero. Therefore, the helicity degree takes the following form:

\begin{equation}
\label{3.3.7}
\theta_N=\frac{Tr[E,O]\prod_{i=1}^{N}\hat{M_i}\left[
\begin{array}{ccc}
O \\
E \\
\end{array}
\right]
}{NTr[E,O]\prod_{i=1}^{N}\hat{M_i}\left[
\begin{array}{ccc}
E \\
O
\end{array}
\right]
}.
\end{equation}

Where E is identity matrix ($\Delta\times\Delta$).

\section{Melting parameters and DMCs}
\label{melting par}
 
\begin{center}
\begin{table}
\resizebox{\textwidth}{!}{\begin{tabular}{ | c | c |  c  | c | c | c | c | c | c | c | c | c | c | c | } 
\hline
\multicolumn{2}{|c|}{N}  & \multicolumn{2}{|c|}{3000} & \multicolumn{2}{|c|}{30000} & \multicolumn{2}{|c|}{60000} & \multicolumn{2}{|c|}{90000} & \multicolumn{2}{|c|}{120000} & \multicolumn{2}{|c|}{180000} \\
\cline{1-2}
  \hline
 \multirow{3}{*}{$T_m$} & $x=0.4$ & 0.217 & 0.211& 0.216 & 0.211& 0.217 & 0.211& 0.217 & 0.211& & & &\\  

  \cline{2-14}& $x=0.5$ & 0.220 & 0.220& 0.220 & 0.219& 0.220 & 0.220& 0.220 & 0.220& & & & \\ 

  \cline{2-14} & $x=0.45$ &  & & 0.218 & 0.214 & 0.218 & 0.215 & 0.218 & 0.215 & 0.218 & 0.216 & 0.218 & 0.216  \\  
  \hline
 \multirow{3}{*}{$\Delta T$} & $x=0.4$ & 0.00212 & 0.00472 & 0.00279& 0.00552& 0.00223& 0.00483& 0.00231& 0.00483& & & &\\  
 \cline{2-14}
  & $x=0.5$ & 0.00194 & 0.00521& 0.00267& 0.00603& 0.00219& 0.00522& 0.00210& 0.00525& & & & \\ 
 \cline{2-14}
 & $x=0.45$ &  & & 0.00468 & 0.01107 & 0.00432 & 0.01053 & 0.00405 & 0.01047 & 0.00408 & 0.01044 & 0.00410 & 0.01038   \\ 
  \hline
\end{tabular}}
{Table 1. Melting temperatures and intervals depending on random or correlated sequences and their lengths.}
\end{table}
\end{center}

\begin{figure}[ht!]
\centering
\includegraphics[width=8.5cm, height=10cm]{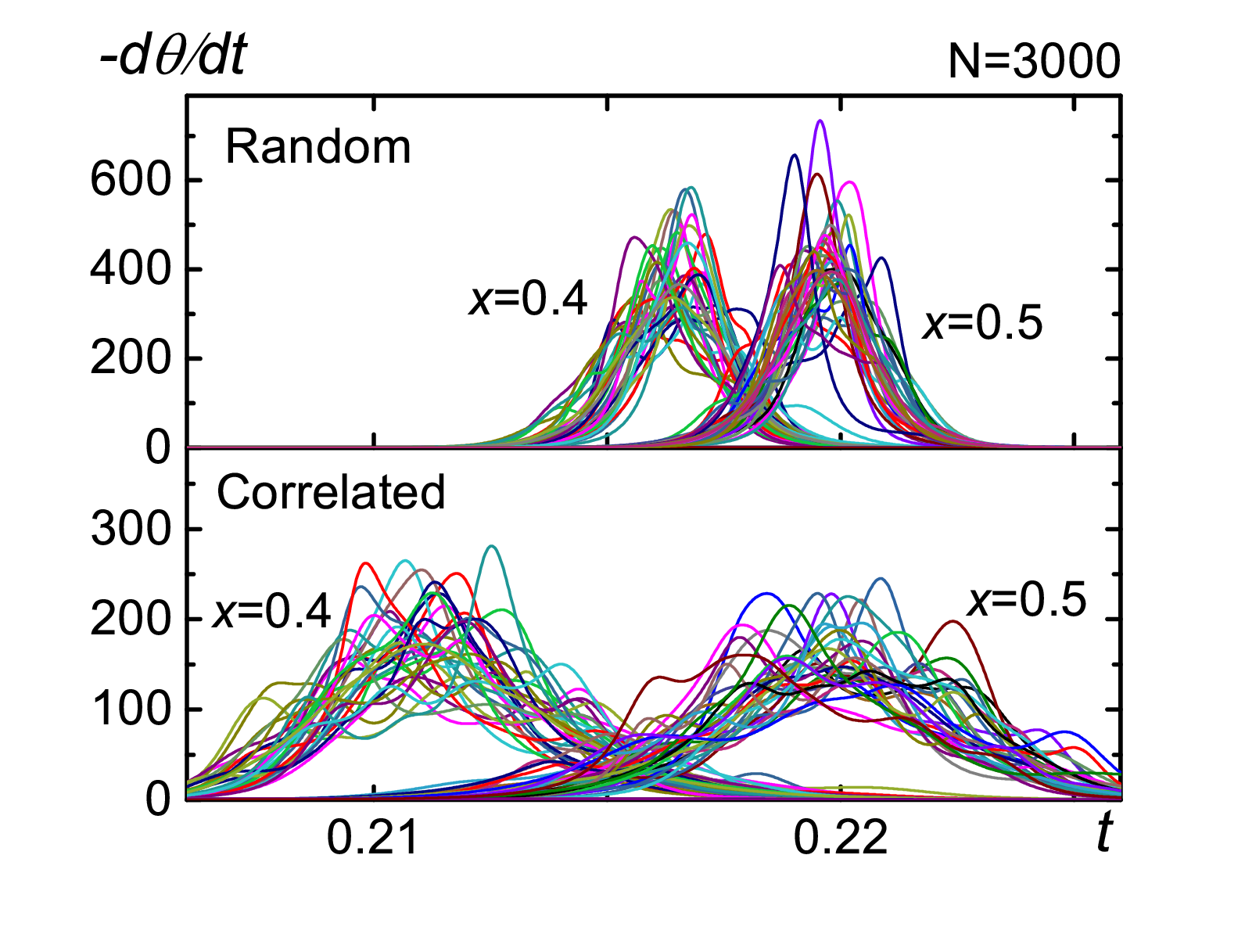}
\caption{\label{all_DMC} DMCs for 30 sequences of each type. }
\end{figure}

\begin{figure}[ht!]
\centering
\includegraphics[width=8.5cm, height=10cm]{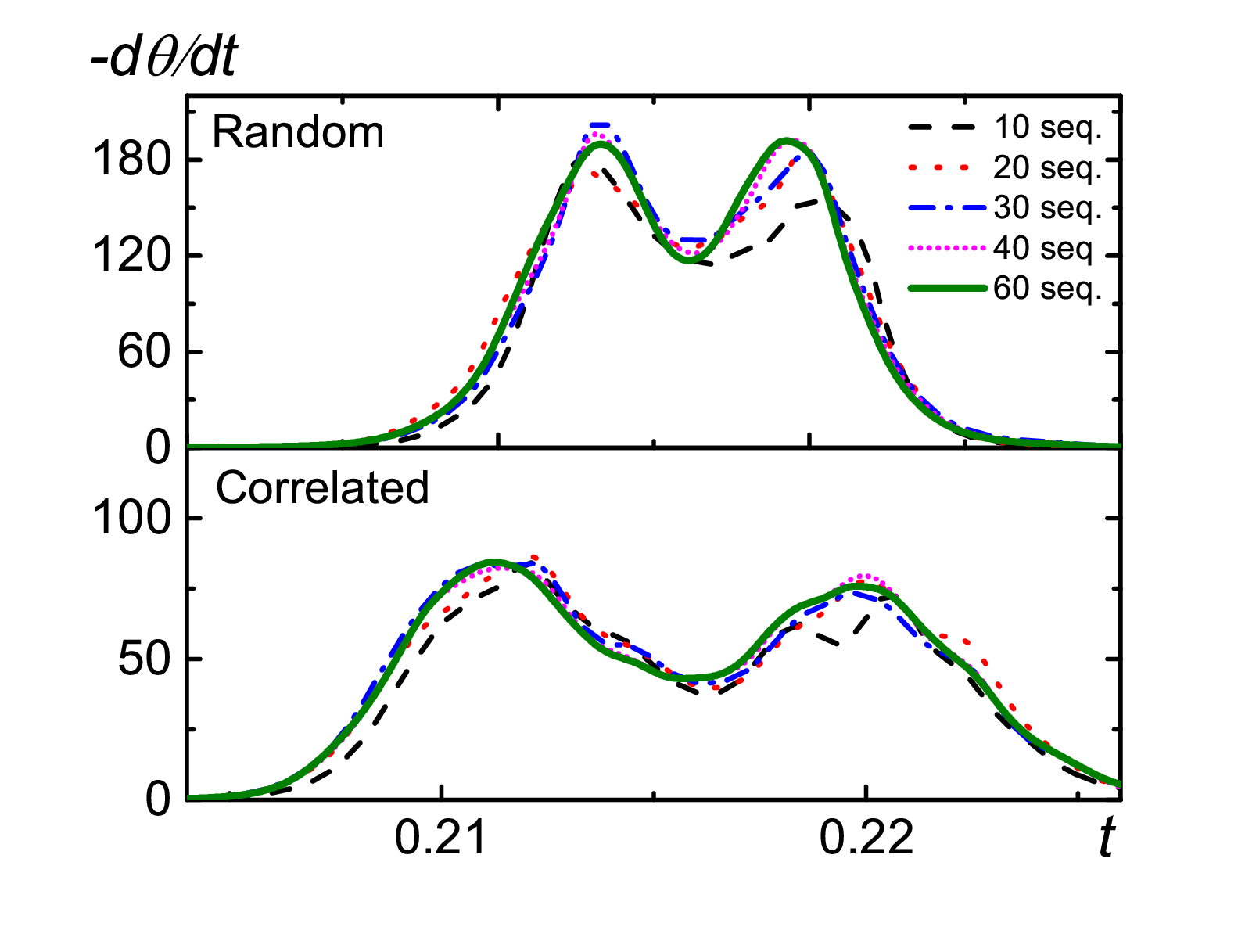}
\caption{\label{all_merged} From 10 to 60 merged sequences, each contains 3000 r.u. As a result block random and block correlated sequences are obtained.}
\end{figure}

From Fig. \ref{all_merged} various length of block systems were obtained through merging of heteropolymers each with 3000 base pairs. The shortest shown contains  30000 r.u. and the longest: 180000 r.u. It is clear, that with increasing number of r.u. the curves become smoother, however, the inclination toward smoothing is more pronounced for random blocks in contrast to correlated blocks.

\pagebreak
\clearpage


%
%

%


 
\bibliographystyle{plain}
\bibliography{Asatryan-etal}

\nocite{*}

\end{document}